\documentclass[%
 reprint,
superscriptaddress,
showpacs,
 amsmath,amssymb,
 aps,
]{revtex4-1}

\usepackage{graphicx}
\usepackage{color}
\usepackage{verbatim}

\begin{document}

\preprint{APS/123-QED}

\title{Comment on "Generating a perfect quantum optical vortex"}

\author{David Barral}
\email{david.barral@c2n.upsaclay.fr}
\author{Jes\'us Li$\tilde{\rm{n}}$ares}
\affiliation{Optics Area, Department of Applied Physics, Faculty of Physics and Faculty of Optics and Optometry, University of Santiago de Compostela, Campus Vida s/n (Campus Universitario Sur), E-15782 Santiago de Compostela, Galicia, Spain.}
\altaffiliation{Present address: Centre de Nanosciences et de Nanotechnologies C2N,
CNRS, Universit\'e Paris Saclay, Marcoussis, France}

\begin{abstract} In a recent article, Banerji {\it et al.} introduced a novel quantum state of light, coined as the perfect quantum optical vortex state [Phys. Rev. A $\bf{94}$, 053838 (2016)] due to its mathematical similarity with the classical perfect vortex beam. This state is obtained by means of the Fourier transform of a Bessel-Gaussian vortex state and the authors claim that this can be accomplished by means of a simple lens. Here, we will show that this statement is wrong since a lens can not modify the quantum noise distribution related to the input optical quantum state and this has to be exchanged by an "effective lens".

\end{abstract}

\maketitle 

A classical perfect vortex beam is that which shows a ring diameter independent of the topological charge. It can be easily obtained by means of the Fourier transform performed by a lens on a Bessel-Gaussian (BG) beam \cite{Vaity2015}. The analogous of this input beam in quantum optics is the BG vortex state, which in cylindrical quantum phase space coordinates at z=0 $(\rho_{0}, \phi_{0})\equiv(\rho, \phi)\vert_{z=0}$, is given by 
\begin{equation}\label{1}
\psi(\rho_{0},\phi_{0})=\frac{i^{q}}{\sqrt{\pi\,I_{q}(\alpha^{2}) }}  \, J_{q} (\sqrt{2} \alpha \rho_{0})\,e^{-(\rho_{0}^{2}-\alpha^{2})/2} \,e^{i q \phi_{0}},
\end{equation}
for $\alpha \in {\rm I\!R}$, as stated in ref. \cite{Banerji2016}. It is important to outline that these abstract phase space coordinates are given by $\rho=\sqrt{\mathcal{E}_{a}^{2}+\mathcal{E}_{b}^{2}}$, $\phi=\arctan(\mathcal{E}_{b}/\mathcal{E}_{a})$, where $\mathcal{E}_{j}$ is the eigenvalue of the optical field-strength operator $\hat{\mathcal{E}}_{j}\propto(\hat{a}_{j}+\hat{a}_{j}^{\dag})$ associated to each mode $j=a,b$ \cite{Schleich2001}. 

However, in contrast, the Fourier transform of this quantum state can not be accomplished by a lens, as claimed in ref. \cite{Banerji2016}, since it only acts on the spatial structure of the beam which supports the quantum state, not on the quantum state itself. In other words, the lens does not act on the optical field-strength space $\mathcal{E}$. Any spatial structure of a two-mode beam, like Gaussian, Hermite-Gaussian, Laguerre-Gaussian and so on, can support a BG quantum vortex. The Fourier transform necessary to obtain the perfect quantum optical vortex state (PQOVS) can be obtained, for instance, by means of the free propagation of the BG vortex state in an homogeneous linear medium where both modes experience the same propagation constant, that is $k_{a}=k_{b}\equiv k$, with $k_{j}=\frac{\omega_{j}}{c} n_{j}$ and $n_{j}$ the refractive index. The quantum optical propagator related to this medium is given by \cite{Linares2012, Barral2015}

\begin{equation}\label{2}
\begin{split}
&K(\rho_{0},\phi_{0},\rho,\phi; 0, z)=\\
&(\frac{i}{2\pi \sin(k z)})^{1/2} \, e^{\frac{-i}{2 \sin(k z)}[(\rho_{0}^{2}+\rho^{2}) \cos(k z) - 2 \rho_{0} \rho \cos(\phi_{0}-\phi)]},\\
\,
\end{split}
\end{equation}
and the quantum state obtained at a plane $z$ of the medium is obtained by means of
\begin{equation}\label{3}
\begin{split}
&\psi(\rho,\phi)=\\
&\int_{0}^{2\pi}\int_{0}^{+\infty} K(\rho_{0},\phi_{0},\rho,\phi; 0, z)\, \psi(\rho_{0},\phi_{0})\, \rho_{0}\,d\rho_{0}\,d\phi_{0}.
\end{split}
\end{equation}
From this, it is then easy to see that the BG quantum vortex given by Equation (\ref{1}) is Fourier transformed at the planes $z=(2m+1)\pi/2k$, with $m$ any integer number. For $m=1$ the following result is obtained
\begin{equation}\label{4}
\psi(\rho,\phi)=\frac{i^{2q+1}}{\sqrt{\pi\,I_{q}(\alpha^{2}) }}  \, I_{q} (\sqrt{2} \alpha \rho) \, e^{-(\rho^{2}+\alpha^{2})/2}\,e^{i q \phi}.
\end{equation}

This is a PQOVS, also known as modified Bessel-Gaussian vortex state (MBG) \cite{Zhu2012}, with core radius $\rho_{c}=\alpha$. 
Besides, it should be outlined that the same result would be obtained in a $z$-inhomogeneous linear medium, but with different Fourier $z$-planes \cite{Barral2015}.

Finally, we would like to point out that the quantum-optics analogous to the outcome obtained by the use of a lens on a Bessel-Gaussian beam could be obtained with a nonlinear medium suitably engineered. To illustrate, two single-mode parametric down-converters with pumps of twice the frequency of the signals and equal gain would produce equal squeezing on both modes of an input state Equation (\ref{1}), generating a radial optical field-strength quantum noise $\langle(\Delta\rho)^{2}\rangle < \langle(\Delta\rho_{0})^{2}\rangle$. In the same way as above, there would be propagation planes where the Fourier transform would be accomplished, but with a scale factor depending on the gain \cite{Linares2012}. In this way the gain would act in an analogous way to the focal length corresponding to the classical case. Therefore, this device would effectively play the role of a lens ("effective lens") in the optical field-strength space $\mathcal{E}$.

\end{document}